\pdfoutput=1

\documentclass{article}
\usepackage[dblblindworkshop,final]{neurips_2026}

\workshoptitle{SocialAgent Workshop}

\usepackage[utf8]{inputenc}
\usepackage[T1]{fontenc}
\usepackage{amsmath,amssymb}
\usepackage{graphicx}
\usepackage{booktabs}
\usepackage{multirow}
\usepackage{xcolor}
\usepackage{hyperref}
\usepackage{url}
\usepackage{subcaption}
\usepackage{microtype}

\title{ODD+LLM: Specifying LLM-Agent Behavior\\for Reproducible Social Simulation}

\author{%
  Justin Downes \\
  Amazon Web Services \\
  \texttt{jusdow@amazon.com} \\
  \And
  Robby Milletich \\
  Amazon Web Services \\
  \texttt{rmilleti@amazon.com} \\
  \And
  Newel Hirst \\
  Amazon Web Services \\
  \texttt{nhirst@amazon.com} \\
}

\begin{document}
\maketitle

\begin{abstract}
The ODD protocol is the de facto standard for describing agent-based models, but it lacks
vocabulary for the components that determine LLM-agent behavior: prompt assembly, memory
serialization, parsing, and retry logic. We propose \textbf{ODD+LLM}, a backward-compatible
extension adding five documentation blocks grounded in the CoALA cognitive-architecture
framework. To show that the gap matters, we conduct a \emph{multiverse analysis}
\citep{steegen2016} of a published LLM naming-game simulation, varying six specification
choices a reimplementer must make. Variants that are defensible readings of the paper stay
within statistical noise of our baseline reconstruction. Missing one instruction that the
paper states but is easy to overlook (that the model answers as an outside observer rather
than as a player) raises the share of runs settling on the convention agents individually
favor from 0.35 to 1.00 and makes convergence three times faster, in a reimplementation
that uses a different model from the original. We complement the demonstration
with a reconstruction-sufficiency review of 28 published LLM-agent simulations, finding that
the details of how prompts are built and how model outputs are parsed are undetermined in
77\% of cases, and that no paper fully specifies its main LLM call.
ODD+LLM gives the field a standard home for information it already produces but currently
scatters across code repositories, supplementary files, and undocumented defaults.
\end{abstract}

\section{Introduction}
\label{sec:intro}

The ODD protocol \citep{grimm2006,grimm2010,grimm2020} standardized agent-based model
description by providing a structured template complete enough that an independent modeler
can reimplement a model and recover its results. Its core is the \textbf{Submodels}
block, where agent behavior is specified as equations or rules; the other blocks describe
the model around it.

LLM-agent simulations have no Submodels block, and no vocabulary for writing one. ODD is
already being adopted for this purpose; YuLan-OneSim \citep{wang2025yulanonesim} generates
executable simulators from LLM-authored ODD protocols across fifty social-science
scenarios, but the protocols it produces specify agent types and interaction patterns while
saying nothing about how context is assembled or how output becomes an action. An LLM
agent's action is the output of a sampling
procedure applied to a token sequence assembled at runtime from a system prompt, a persona,
some subset of interaction history, some rendering of environment state, and an output
instruction, then parsed, possibly rejected, possibly resampled, and mapped onto a
simulation action. That assembly procedure is what a Submodels block for LLM agents must
describe.

This matters in practice. \citet{ye2026} hold persona \emph{content} fixed and vary only its
\emph{presentation} (prose, bullet list, key--value table), observing a
76-percentage-point difference in cooperation rate ($N{=}30$ seeds, $p{<}0.001$). Two papers
reporting ``agents were given demographic personas'' can therefore support opposite
conclusions about whether LLM agents cooperate.

ODD+LLM is a reporting standard, not a new algorithm. Like ODD itself and reporting
guidelines such as CONSORT \citep{schulz2010consort} and TRIPOD \citep{collins2015tripod},
its contribution is what it standardizes and the evidence that those items matter. We make
three contributions:
\begin{enumerate}
  \item \textbf{ODD+LLM}: a backward-compatible extension to ODD adding five documentation
    blocks for LLM-agent simulations (\S\ref{sec:protocol}). Blocks~C and D use
    \emph{repeated records} (one per model role and one per call site), following ODD's
    existing pattern for submodels. The core of the protocol is Block~D, the behavioral submodel,
    which specifies each LLM call in the loop by signature, binding, and resolution.
  \item \textbf{A multiverse demonstration} that specification salience is consequential
    (\S\ref{sec:multiverse}). We take a published naming-game simulation
    \citep{ashery2025}, identify six specification choices (four from supplementary
    materials, one in main text but non-obvious, one genuinely open), implement a baseline
    and seven one-factor variants, and show that missing a single stated but easily
    overlooked instruction determines the headline result.
  \item \textbf{A reconstruction-sufficiency review} of 28 LLM-agent simulations coded for
    where an independent implementer would have to guess (\S\ref{sec:review}).
\end{enumerate}

\section{Related Work}
\label{sec:related}

\paragraph{ODD and its extensions.}
ODD \citep{grimm2006,grimm2010} was extended to ODD+D for decision-making
\citep{muller2013}, ODD+2D for data-driven models \citep{laatabi2018}, vODD-DD for visual
documentation \citep{lee2026vodddd}, and VISA for machine-reproducible descriptions
\citep{he2026visa}. None address the prompt-assembly problem because it did not exist
before LLM agents. The closest antecedent is the move from verbal rules to equations in ODD
2020 \citep{grimm2020}, which argued that natural-language descriptions of behavior are
insufficient for reconstruction, precisely the claim we make about LLM-agent descriptions.
VISA opens with nearly our framing (``two readers routinely reconstruct different models
from the same documentation'') but targets classical ABMs rather than LLM-agent behavior.

\paragraph{ODD applied to LLM-agent simulation.}
YuLan-OneSim \citep{wang2025yulanonesim} generates executable LLM-agent simulators from
ODD protocols across 50 social-science scenarios. They use ODD
\emph{forward} (natural language $\to$ ODD $\to$ code), whereas we test it \emph{backward} (published study $\to$ could an implementer reconstruct it?). Their
generated protocols specify agent types and interaction patterns but contain nothing about
which model runs, how context is assembled, or how output is parsed; those choices are
resolved downstream by code templates and configuration. A system that already wants ODD for LLM agents is evidence of demand; that its protocols
stop where Blocks~C and~D begin is evidence of the gap.

\paragraph{Validity critiques of LLM-agent simulations.}
\citet{larooij2025} review 35 papers and find systematic validation gaps.
\citet{zhou2025} propose six design-validity principles (PIMMUR) and re-run five papers.
\citet{ye2026} demonstrate that format-only perturbations move outcomes by up to
76pp, though the same perturbation on deepseek-v3 produces ${\sim}$1pp, indicating strong
model-dependence. \citet{barrie2025} argue that naming-game results are observationally
equivalent to data leakage from pretraining; \citet{ashery2025reply} respond that the
effect survives controls. Our work complements these critiques by providing a
\emph{specification}, and our multiverse analysis measures the range of outcomes
consistent with a description rather than testing whether a specific mechanism is correct.

\paragraph{Representation-effect audits.}
\citet{ye2026}'s TRAILS-R taxonomy audits representation-level perturbations (formatting,
delimiter choice, output schema, instruction order, memory representation) that map onto our
Block~D items DB-1 through DB-7. We adopt their three-level \texttt{claim\_strength}
hierarchy (exploratory, mechanism, policy) for our corpus metadata. TRAILS is an
\emph{audit} taxonomy (what to perturb to test robustness), while ODD+LLM is a
\emph{documentation} protocol (what to write down so a
finding can be reconstructed at all). Block~D's resolution layer (parser, validity
constraints, retry, failure cascade, failure rate) has no TRAILS analogue, and our
reconstruction-sufficiency review measures determinability rather than robustness.
TRAILS found that memory representation had the smallest effect among the representation
axes they tested, independently corroborating our Stage~2 finding that memory-format
and memory-order arms (M1, M2) produce small population-level effects relative to person
framing (M4).

\paragraph{Cognitive-architecture frameworks.}
CoALA \citep{sumers2024} provides a taxonomy of LLM-agent components (working memory,
long-term memory, internal/external actions, decision procedure). We adopt it as the
ontology for Block~A rather than inventing one. DSPy \citep{khattab2024} provides a
machine-readable signature format for LLM modules; Block~D's schema draws on it.

\section{The ODD+LLM Protocol}
\label{sec:protocol}

ODD+LLM adds five blocks to the standard ODD template. The existing ODD blocks (Purpose,
Entities/State Variables, Process Overview, Design Concepts, Initialization, Input Data,
Submodels) are retained unchanged; the extension occupies a new section between Design
Concepts and Initialization.

Following ODD's own pattern (Submodels is already one entry per submodel; State Variables is
a list), \textbf{Blocks C and D are repeated records}. Block~C has one record per model role
and Block~D one per LLM call site, so documentation grows with the system. A two-agent
dilemma with one call site stays a one-page specification, while a generative-architecture paper with nine distinct calls produces nine
records.

\subsection{Block A: Cognitive Architecture (CoALA-structured)}

Documents the agent's information-processing architecture using CoALA's taxonomy (9 items):
working memory contents per cycle; long-term memory types, backing store, retrieval rule
(formula and cardinality), and write rule; internal and external action spaces; decision
procedure (including actions per activation); tools; and the inter-agent communication
channel with its visibility parameter.

\subsection{Block B: Identity}

Documents persona source, content, surface form (the verbatim template), heterogeneity
method, behavioral constraints, and identity stability (static vs.\ drifting, with the
update rule). The surface-form item is marked as behavioral rather than cosmetic, following
\citet{ye2026}'s finding. (6 items.)

\subsection{Block C: Inference Configuration}
\label{sec:blockc}

Block~C has a two-item \textbf{spine} at the system level and a \textbf{per-role record}
(6 items) repeated for each model or API in the behavior or measurement path.

\paragraph{C-spine (system level).}
\begin{itemize}
  \item \textbf{CS1}: Known provider-side non-determinism sources (batching, routing,
    caching) and their handling.
  \item \textbf{CS2}: Execution concurrency and ordering: whether calls are issued
    sequentially, concurrently, or asynchronously, and whether intra-timestep ordering is
    deterministic.
\end{itemize}

\paragraph{C-role (repeated per model role).}
Roles observed in the corpus include \emph{agent}, \emph{classifier}, \emph{judge},
\emph{embedder}, \emph{generator} (offline content), and \emph{external API}. Each record
contains:
\begin{itemize}
  \item \textbf{CR1--CR3}: Model identity (exact version string or pinned snapshot),
    build identity (system fingerprint), and provider/endpoint/API version/access date.
  \item \textbf{CR4}: Decoding parameters, named unambiguously and distinguished from
    same-named training or reward parameters.
  \item \textbf{CR5}: Serving precision and quantization.
  \item \textbf{CR6}: Training or adaptation (base checkpoint, method, data construction,
    objective, selection criterion). Marked NA if used off the shelf.
\end{itemize}

Every model role gets a record. An unnamed embedder is a role record scoring~0 on CR1, not
a footnote.

\subsection{Block D: Behavioral Submodel}
\label{sec:blockd}

Block~D is the ODD Submodels equivalent for LLM agents. Like Block~C, it has a two-item
\textbf{spine} at the system level and a \textbf{per-call-site record} (15 items) repeated
for each distinct LLM call in the simulation loop. Items within a call-site record are
grouped into four sections:

\paragraph{D-spine (system level).}
\begin{itemize}
  \item \textbf{DG1}: Temporal coupling: whether step $t$ output enters step $t{+}1$
    input, and in what form.
  \item \textbf{DG2}: Scheduling and activation: timestep duration, who acts when,
    within-step ordering. Controlled vocabulary adopted from Concordia
    \citep{vezhnevets2023}:
    \texttt{simultaneous}/\texttt{sequential}/\texttt{asynchronous}.
\end{itemize}

\paragraph{Signature (what the call is).}
\begin{itemize}
  \item \textbf{DS-1}: Input bindings: which state variables feed this call.
  \item \textbf{DS-2}: Output type or schema.
\end{itemize}

\paragraph{Binding (how it is rendered).}
\begin{itemize}
  \item \textbf{DB-1}: Static text (system prompt and fixed instructions, verbatim).
  \item \textbf{DB-2}: Serialization per slot: how each variable renders into text.
  \item \textbf{DB-3}: Component order and system-vs-user role placement.
  \item \textbf{DB-4}: History policy with cardinality (``top-ranked memories'' without a
    $k$ is underdetermined).
  \item \textbf{DB-5}: Truncation/summarization rule with constants.
  \item \textbf{DB-6}: Token budget and drop order.
  \item \textbf{DB-7}: Action-space encoding: how the option set is presented.
\end{itemize}

\paragraph{Resolution (how output becomes an action).}
\begin{itemize}
  \item \textbf{DR-1}: Parser: exact extraction procedure.
  \item \textbf{DR-2}: Validity constraints: what makes a parseable output invalid.
  \item \textbf{DR-3}: Retry policy: count, whether prompt is modified, fallback on final
    failure.
  \item \textbf{DR-4}: Failure action and cascade: what the system does, and what other
    agents observe.
  \item \textbf{DR-5}: Failure rate (an empirical result, included to force reporting;
    analogous to ODD's ``Output Data'').
\end{itemize}

\paragraph{Delegation.}
\begin{itemize}
  \item \textbf{DF-1}: Third-party framework: library name, \emph{pinned version}, and
    which part of assembly or resolution it owns. An unpinned framework dependency is
    undetermined.
\end{itemize}

\paragraph{Systems with multiple call sites.}
Complex architectures may call the LLM in several places (e.g., acting, reflecting,
summarizing memory). Document the \emph{primary} call site, the one that produces the
agent's behavioral output, against all 15 items. List any other call sites in a short
inventory that names each one and marks whether it is fully documented. A simple model
stays at one record, and readers can still see how much of a complex architecture is
specified.

Block~D is designed so that a \texttt{.yaml} configuration file satisfying the schema
\emph{is} the Block~D documentation; no separate prose description is needed. The
arm configurations for our multiverse analysis (\S\ref{sec:multiverse}) are themselves
Block~D specifications.

\subsection{Block E: Validation}

Documents behavioral reference, validation level, replication count (E3a) and the unit over
which variance is reported (E3b), variance reporting, representation sensitivity (whether
any prompt-form perturbation was tested), population dispersion, ablation and non-social
controls (E7a), non-LLM or human baselines (E7b), and longitudinal stability. (10 items.)

The E3 and E7 splits distinguish papers that state $n{=}1$ from those that never state $n$,
and papers with ablation controls from those with human baselines, distinctions that
collapsed in an earlier draft and obscured the dominant failure mode.

\section{Multiverse Demonstration}
\label{sec:multiverse}

We select \citet{ashery2025}, ``Emergent social conventions and collective bias in LLM
populations'' (\emph{Science Advances}, 2025). In this naming game, $N{=}24$ agents
repeatedly choose names from a pool of $W{=}2$ letters; matching earns $+100$, mismatching
costs $-50$. The paper reports that individually unbiased agents converge
disproportionately to one ``strong'' convention, driven by microscopic asymmetry in
stick-after-success vs.\ switch-after-failure behavior.

We chose this paper because it is well documented by current norms. It reports model
versions, includes real controls, provides supplementary prompts, and publishes code
(\texttt{github.com/Ariel-Flint-Ashery/AI-norms}). This makes it a conservative test case.
The gaps we identify exist despite above-average documentation, because there is no
standard home for the information Block~D requires.

\paragraph{Branch points.}
We identify six specification choices from the 21 decision points our instrument assigns to
this paper, selected on the basis that each corresponds to a choice our rubric flags as
behaviorally relevant (\texttt{changes\_behavior: certain} or \texttt{likely}):

\begin{table}[h]
\centering
\small
\begin{tabular}{llp{4.2cm}ll}
\toprule
ID & Axis & Main-text ambiguity & Levels & SI \\
\midrule
M1 & Memory format & ``dynamic memory prompt'' & table / narr.\ / JSON & SI \\
M2 & Memory order & Order of $H$ entries unstated & recent / oldest-first & SI \\
M3 & Score semantics & ``accumulated score over $H$'' & window / cumulative & SI \\
M4 & Person framing & ``external observer'' & 3rd / 1st-person & Spec. \\
M5 & Pairing protocol & ``two agents randomly selected'' & random / perfect & SI \\
M6 & Parse failure & ``consistent format'' & resample+uniform / uniform-only & Open \\
\bottomrule
\end{tabular}
\caption{Branch points (multiverse axes). SI = determinable from the supplementary
materials; Spec.\ = specified in the main text but non-obvious; Open = genuinely
underspecified. Four of six choices are resolvable from supplementary materials, one is
specified in the main text but non-obvious, and a reader working from the methods section
alone would likely resolve them differently.}
\label{tab:branches}
\end{table}

We implement eight arms: one baseline (our best reading of the main text) plus one
alternative per branch point (with M1 having two alternatives). All arms share $N{=}24$,
$W{=}2$, $H{=}5$, payoffs $+100/-50$, and chain-of-thought prompting, everything the paper
\emph{does} specify. We use Claude Sonnet 4.6 (via AWS Bedrock, temperature 0.7) rather
than the paper's original models (which specify temperature 0.5, top-k 10, max tokens 6 in
Table~S4 of the supplementary). Both deviations illustrate the specification-salience problem,
because the original values are in the supplementary materials but are easy to miss.

\paragraph{Stage 1: Memory-state probe.}
Before running full simulations, we synthesize memory states and query the model directly
(7 states $\times$ 8 arms $\times$ 300 samples = 16,800 calls). This measures the
microscopic asymmetry the paper identifies as the origin of collective bias.

\begin{table}[h]
\centering
\small
\begin{tabular}{lcc}
\toprule
Arm & $P(M \mid \text{empty})$ & R4 Asymmetry \\
\midrule
Baseline (table/recent/3rd) & 0.680 & $+0.200$ \\
M1: Narrative & 0.653 & $+0.063$ \\
M1: JSON & 0.730 & $-0.040$ \\
M2: Oldest-first & 0.690 & $+0.021$ \\
M3: Cumulative score & 0.707 & $+0.217$ \\
\textbf{M4: First-person} & \textbf{0.587} & $\mathbf{-0.243}$ \\
M5: Perfect matching & 0.693 & $+0.257$ \\
M6: Uniform fallback & 0.663 & $+0.227$ \\
\bottomrule
\end{tabular}
\caption{Stage 1 probe results. R4 asymmetry measures the difference
$P(M \mid \{1{:}M,Q;\; 2{:}Q,M\}) - P(Q \mid \{1{:}Q,M;\; 2{:}M,Q\})$ (matching
Flint~Ashery et al.'s interaction indexing): positive means M is the ``strong''
convention. M4 (person framing) reverses the sign; M1 (format) and M2 (order) attenuate
or reverse it. M5 and M6 cannot logically affect a direct probe (they are
population-level mechanisms); their variation is sampling noise.}
\label{tab:probe}
\end{table}

\paragraph{Stage 2: Full simulation.}
We run 20 seeds per arm (8 arms $\times$ 20 seeds $\times$ $\sim$200 calls/seed
$\approx$ 32,000 LLM calls). The headline metric is $P(M)$: the fraction of runs whose
consensus convention is M.

\begin{table}[h]
\centering
\small
\begin{tabular}{lccc}
\toprule
Arm & $P(M)$ & 95\% Wilson CI & Conv.\ round \\
\midrule
Baseline & 0.350 & [0.18, 0.57] & 16.1 \\
M1: Narrative & 0.450 & [0.26, 0.66] & 15.6 \\
M1: JSON & 0.300 & [0.15, 0.52] & 17.6 \\
M2: Oldest-first & 0.450 & [0.26, 0.66] & 19.9 \\
M3: Cumulative & 0.350 & [0.18, 0.57] & 14.9 \\
\textbf{M4: First-person} & \textbf{1.000} & \textbf{[0.84, 1.00]} & \textbf{5.3} \\
M5: Perfect matching & 0.550 & [0.34, 0.74] & 19.4 \\
M6: Uniform fallback & 0.350 & [0.18, 0.57] & 15.3 \\
\bottomrule
\end{tabular}
\caption{Stage 2 results. Conv.\ round = first population round at which mean success
rate crosses the 0.95 threshold; arms may subsequently dip below it (see Figure~2).
First-person framing (M4) produces convergence to M in all 20 seeds,
three times faster than any third-person arm.}
\label{tab:stage2}
\end{table}

\begin{figure}[t]
\centering
\includegraphics[width=0.95\textwidth]{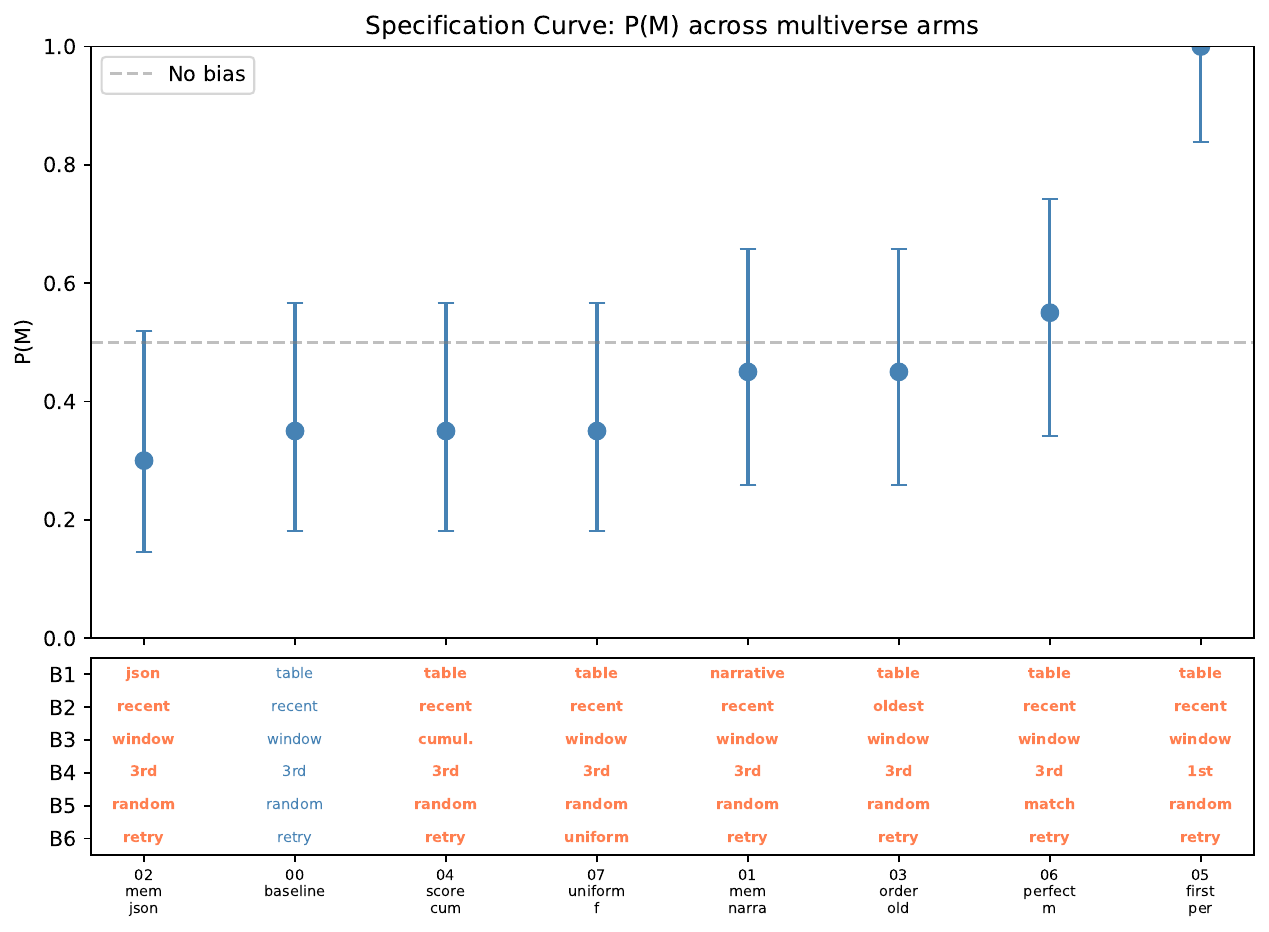}
\caption{Specification curve \citep{simonsohn2020}: $P(M)$ across multiverse arms
(sorted), with Wilson CIs. The dot matrix below shows which branch level each arm uses.
The first-person arm (M4, rightmost) is the only arm whose CI does not overlap the
others; a single specification choice qualitatively determines the outcome.}
\label{fig:spec_curve}
\end{figure}

\begin{figure}[!ht]
\centering
\includegraphics[width=0.85\textwidth]{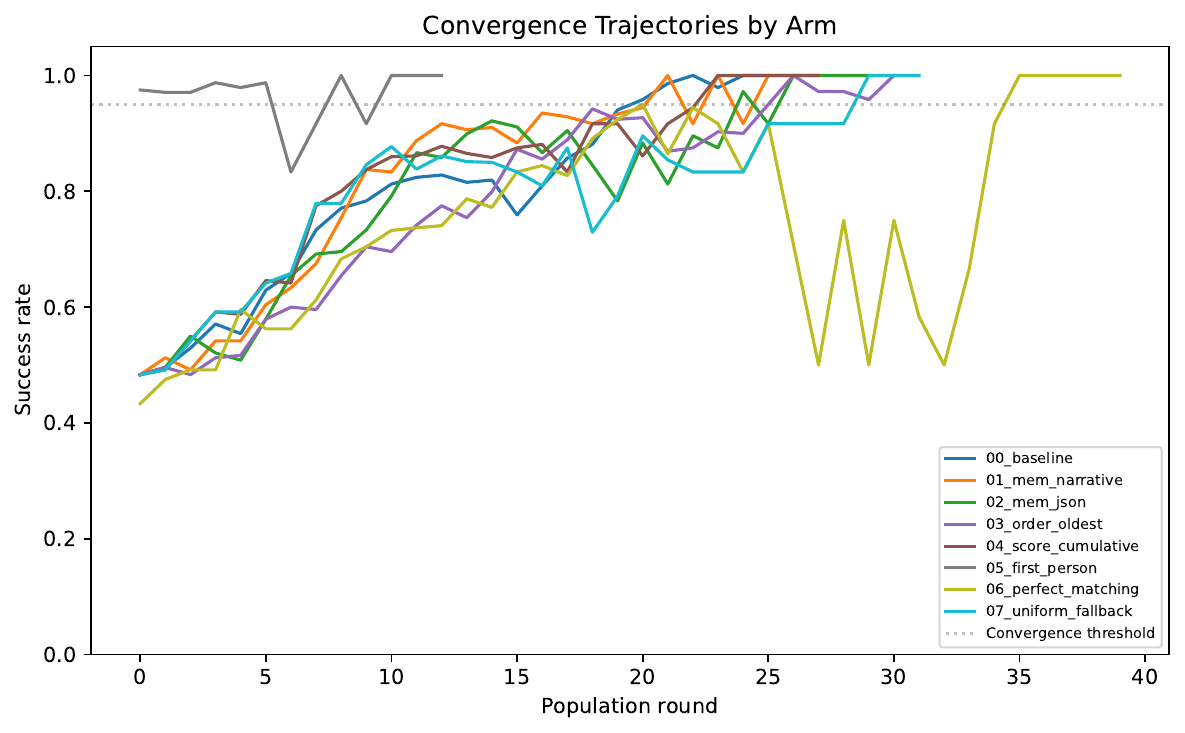}
\caption{Mean success rate by population round. First-person framing converges by round~5;
most third-person arms converge by rounds 15--20, though one arm (M2: oldest-first) shows
a slower trajectory reaching full coordination only near round~35. All arms produce
naming-game-like convergence (they are all plausible implementations), but they land on
different conventions.}
\label{fig:convergence}
\end{figure}

\paragraph{Interpretation.}
The specification curve \citep{simonsohn2020} (Figure~\ref{fig:spec_curve}) separates two
kinds of arm. The six arms that vary a choice the main text leaves open or delegates to the
supplement span $P(M) \in [0.30, 0.55]$, all within noise of the baseline. M4 is different
in kind. It reverses an instruction the paper states, so it tests what happens when salient
information is missed rather than offering a seventh defensible reconstruction. Two tensions constrain the claim.

\textbf{Stage~1 vs.\ Stage~2.} The microscopic probe (Table~\ref{tab:probe}) shows M as
the ``strong'' convention (baseline asymmetry $+0.200$), yet the population simulation
lands on Q two-thirds of the time ($P(M){=}0.350$). This reversal likely reflects
stochastic dynamics at 20~seeds. Every third-person arm's Wilson CI covers 0.5, so none
can be distinguished from a fair coin. The detectable signal is M4 alone.

\textbf{A single-item claim.} We do not claim that six choices determine the outcome.
Missing \emph{one} stated instruction, person framing, changes it in our reimplementation ($P(M){=}1.00$ in all 20~seeds, convergence three times faster;
Figure~\ref{fig:convergence}), while five further choices plausibly modulate the outcome
below our detection power. Because we also substitute the model and decoding settings, the
result shows sensitivity in this reimplementation, not a property of the original system.
It still matters in practice, because an implementer who writes first-person instructions (the
natural default for ``a naming-game simulation'') would observe a qualitatively different
phenomenon.

\textbf{Specification salience.} M4 is specified in the paper (``external observer
forecasting Player~1'') and is thus not underspecified in the strict sense. But it
demonstrates that even specified information can fail to reach an independent reader if it
has no standard home. Only M6 (parse failure) is genuinely open, and it returns
the baseline $P(M)$ of 0.350 exactly. Parse failures occur in under 1\% of calls, so
the branch almost never fires, which shows where specification effort matters.

\section{Reconstruction-Sufficiency Review}
\label{sec:review}

We code 28 published LLM-agent simulations (2023--2026) for reconstruction sufficiency,
stratified by architecture: minimal-choice agents, opinion dynamics, platform-scale
simulators, economic/market models, full generative architectures, policy testbeds, and
norm emergence. For each paper, we ask one question per item: \emph{could an independent
implementer, reading only this paper and its linked artifacts, determine this, or would
they have to guess?}

\paragraph{Instrument.} Items are scored 0 (undetermined), 1 (partially determined: the
component is named but at least one behaviorally relevant aspect remains open), or 2
(determined). The key distinction is between asserting that a component exists and specifying it. ``Agents have a
memory module'' asserts a component exists; it does not specify what enters it, how it is
rendered, or when it is cleared. Assertion without operational content scores~1, never~2.
Every score of 0 or 1 generates a named decision point, an independently variable choice
that two implementers could set differently without contradicting the text.

The instrument was derived from 10 papers screened in Pass~1 (9 eligible, 1 excluded on
read), during which the repeated-record structure emerged from corpus evidence
then frozen as v1. Pass~2 codes the remaining 19 papers with
the frozen instrument unchanged; decision points that do not fit the block structure are
recorded as misses and reported.

Each paper carries a \texttt{claim\_strength} label (exploratory, mechanism, or policy;
following the TRAILS hierarchy \citep{ye2026}), letting us test whether the strongest
empirical claims rest on the weakest specifications.

\paragraph{Results.}
Of 32 papers screened, 4 were excluded on read (2$\times$E1, 1$\times$E3, 1 deferred), yielding 28
coded papers across 18 venues and 7 strata (2023--2026); all 28 are listed individually in
Table~\ref{tab:perpaper}. Because Blocks~C and D are repeated records, papers have
different denominators and raw item counts are not comparable. We therefore report three
numbers.

\textbf{(1) System-block coverage.} The mean sufficiency score across the 29 system-level
items is 1.52/2.00 (76\%). Architecture (Block~A: 1.63) and validation (Block~E: 1.46) are
documented reasonably well. Inference configuration is not. Block~C averages 0.48/2.00
per role and 0.00 on system fingerprint (CR2), which scored~0 in all 28 papers.

\textbf{(2) Fraction of papers whose primary call site is fully determined.} None of
the 28 primary call sites scored~2 on every binding and resolution item. The resolution
layer (DR$_1$--DR$_5$) is nearly absent from the literature: 96\% of resolution
item-scores are $<$2 (i.e., not fully determined); retry policy (DR$_3$) scores~0 in
85\% of papers; parse-failure rate (DR$_5$) is unreported in 93\%. Block~D binding items
(DB$_1$--DB$_7$) fare better at 1.14/2.00 but still leave 62\% of item-scores at 0 or~1.
The combined undetermination rate across binding and resolution items (fraction of
item-scores $<$2) is \textbf{77\%} (236/306).

\textbf{(3) Complexity.} Papers average 1.5 model roles (range 1--3); only the primary
call site is formally coded. Block~C averages 0.53/2.00 for agent roles alone but 0.48
across all roles, confirming that secondary models are systematically invisible.

\begin{table}[h]
\centering\small
\begin{tabular}{lcccc}
\toprule
Block & Scope & Mean & Items at 0 & $n$ \\
\midrule
A (Architecture) & system & 1.63 & n/a & 209 \\
B (Identity) & system & 1.57 & n/a & 131 \\
C spine & system & 0.90 & n/a & 48 \\
D spine & system & 1.73 & n/a & 56 \\
E (Validation) & system & 1.46 & E3a (37\%) & 269 \\
\midrule
C per-role & repeated & 0.48 & CR2 (100\%) & 186 \\
D call-site & repeated & 0.90 & DR$_5$ (93\%) & 370 \\
\quad Binding & & 1.14 & & 170 \\
\quad Resolution & & 0.37 & & 136 \\
\bottomrule
\end{tabular}
\caption{Mean sufficiency by block (0--2 scale). System-level items that existing
documentation conventions cover (A, D-spine: mean 1.68) score roughly double the
repeated-record items (C per-role, D call-site: mean 0.69) that have no standard home.
$n < 28 \times \text{items}$ because NA items (design makes the item inapplicable) and
UNVERIFIED items (supplementary not retrieved) are excluded from scoring.}
\label{tab:coverage}
\end{table}

\paragraph{Decision points.} The 28 papers yield 705 decision points (mean 25.2 per paper,
range 10--40). The inventory saturated by paper~15: fewer than 2 novel axes per paper
thereafter, suggesting that the block structure derived from Pass~1 generalizes to the
held-out set. Zero decision points required a block the frozen instrument could not house.

\paragraph{Claim strength vs.\ specification.} Papers making mechanism claims ($n{=}11$)
average a per-paper Block~D call-site score of 1.00/2.00; policy claims ($n{=}9$) average
0.72. Stratum is a confounder (policy papers cluster in more complex architectures) and
the difference is not statistically tested at this sample size.

\section{Discussion}
\label{sec:discussion}

\paragraph{The protocol records choices already made.}
Every Block~D item corresponds to a choice the implementer has already made; the protocol
asks only that it be written in a standard location. For our own study this took one
configuration file plus a results section (DR-5, Block~E; Appendix~\ref{app:worked}). We
have not measured author burden more generally. Our multiverse arm configurations
\emph{are} Block~D specifications, and losing one stated item from them produced a
qualitatively different outcome.

\paragraph{Limitations.}
\emph{Demonstration.} The multiverse is a single case study with 20 seeds per arm. It
shows that one specification item can determine a result but does not estimate how often
this happens across simulations. It uses one model (Claude Sonnet 4.6) with decoding
settings that differ from the original, so magnitudes may differ with other models. The
design is one-factor-at-a-time (8 arms), not a full factorial ($2^6{=}64$); interactions
between axes are untested. With 20 seeds, only M4 is statistically separable; the remaining
branches may modulate outcomes below our detection power. M4 is specified in the source
paper (``external observer''), so first-person framing is a sensitivity test of what happens
when salient-but-non-obvious information is missed, not a defensible reconstruction.
\emph{Coverage of the protocol.} The multiverse exercises six items (DB-1 to DB-4, DG2,
DR-3). The remaining items are motivated by how often the corpus leaves them undetermined
and by prior evidence such as \citet{ye2026}, not by experiment; validating each item is
future work. \emph{Evidence for the standard.} The rubric is new and was applied by a
single coder without inter-rater reliability estimation, so the 77\% figure is indicative
rather than validated. We have not yet shown that ODD+LLM improves reconstruction or
measured its cost to authors. Double-coding a subset of the corpus and a study in which
readers reconstruct calls from completed specifications are the next steps. The current
protocol assumes text-in/text-out calls; multimodal inputs (images, video frames) to
vision-language models are not covered.

\paragraph{Not a critique.}
Flint~Ashery et al.\ is well-documented by current norms. The information exists;
what is missing is \emph{salience}, a standard place for choices that are already made.

\paragraph{Conclusion.}
ODD+LLM gives the LLM-agent simulation community a structured home for specification
decisions that determine behavior. It is a documentation convention that reduces avoidable
ambiguity. It cannot guarantee reproducibility, because models change, providers update silently, and
stochastic systems are inherently variable. But the choices a researcher \emph{can} pin down
should be pinned down. Missing one stated instruction flips a reimplementation's headline
result; 77\% of binding and resolution items across 28 papers are undetermined. Adopting the protocol
costs one configuration file; Figure~\ref{fig:spec_curve} shows the cost of leaving these
choices undocumented.

\bibliographystyle{plainnat}
\bibliography{references}

\newpage
\appendix
\section{Per-Paper Sufficiency Scores}
\label{app:scores}

Table~\ref{tab:perpaper} reports the sufficiency scores for all 28 coded papers. System
mean is over the 29 system-level items (Blocks A, B, C-spine, D-spine, E). D-site mean
is over the 15 per-call-site items for the primary call site. DPs is the count of
decision points, independently variable choices the text leaves open.

\begin{table}[h]
\centering\small
\begin{tabular}{llcccccl}
\toprule
Paper & Str. & Year & Sys & D-site & DPs & Roles & Claim \\
\midrule
\citealt{akata2025}      & S1 & 2025 & 1.86 & 1.73 & 10 & 1 & mechanism \\
\citealt{ashery2025}     & S1 & 2025 & 1.64 & 0.64 & 21 & 1 & mechanism \\
\citealt{fontana2025}    & S1 & 2025 & 1.89 & 1.17 & 13 & 1 & n/a \\
\citealt{li2025}         & S1 & 2025 & 1.77 & 1.29 & 14 & 1 & mechanism \\
\citealt{piatti2024}     & S1 & 2024 & 1.78 & 0.93 & 20 & 2 & mechanism \\
\midrule
\citealt{chuang2024}     & S2 & 2024 & 1.69 & 1.23 & 17 & 1 & n/a \\
\citealt{gu2025}         & S2 & 2025 & 1.27 & 0.36 & 32 & 2 & mechanism \\
\citealt{mou2024}        & S2 & 2024 & 1.67 & 1.08 & 20 & 1 & n/a \\
\citealt{piao2025a}      & S2 & 2025 & 1.22 & 0.64 & 28 & 1 & policy \\
\citealt{wang2025a}      & S2 & 2025 & 1.33 & 0.55 & 30 & 1 & policy \\
\midrule
\citealt{guan2025}       & S3 & 2025 & 1.79 & 1.07 & 26 & 3 & mechanism \\
\citealt{piao2025b}      & S3 & 2025 & 1.11 & 0.47 & 26 & 1 & policy \\
\citealt{tornberg2023}   & S3 & 2023 & 1.32 & 0.71 & 32 & 2 & policy \\
\citealt{yang2025a}      & S3 & 2025 & 1.24 & 0.12 & 26 & 1 & n/a \\
\midrule
\citealt{manning2024}    & S4 & 2024 & 1.74 & 1.21 & 24 & 2 & mechanism \\
\citealt{nianli2024}     & S4 & 2024 & 1.88 & 1.15 & 15 & 1 & n/a \\
\citealt{wang2025c}      & S4 & 2025 & 1.36 & 1.00 & 33 & 3 & mechanism \\
\citealt{yang2025b}      & S4 & 2025 & 1.38 & 1.00 & 31 & 2 & mechanism \\
\midrule
\citealt{dai2024}        & S5 & 2024 & 1.52 & 1.07 & 28 & 1 & mechanism \\
\citealt{gao2023}        & S5 & 2023 & 1.04 & 0.33 & 40 & 2 & policy \\
\citealt{zhilinwang2023} & S5 & 2023 & 1.36 & 0.73 & 37 & 2 & mechanism \\
\midrule
\citealt{aoki2026}       & S6 & 2026 & 1.81 & 1.13 & 16 & 1 & policy \\
\citealt{hou2025}        & S6 & 2025 & 1.78 & 1.08 & 20 & 1 & n/a \\
\citealt{huang2026}      & S6 & 2026 & 1.54 & 0.56 & 26 & 1 & n/a \\
\citealt{liu2025}        & S6 & 2025 & 1.46 & 1.13 & 33 & 3 & policy \\
\midrule
\citealt{liu2024}        & S7 & 2024 & 1.44 & 1.00 & 25 & 1 & policy \\
\citealt{orlando2026}    & S7 & 2026 & 1.50 & 0.53 & 36 & 3 & policy \\
\citealt{ren2024}        & S7 & 2024 & 1.23 & 0.89 & 26 & 1 & n/a \\
\bottomrule
\end{tabular}
\caption{Per-paper sufficiency scores (0--2 scale). Strata: S1 minimal-choice, S2 opinion
dynamics, S3 platform-scale, S4 economic/market, S5 generative architectures, S6 policy
testbeds, S7 norms/emergence. Claim: mechanism = mechanistic causal claim; policy =
policy-relevant conclusion; n/a = exploratory or unclassified.
Wang et al.\ 2025a (COLING) and 2025c (ACM TOIS) are corpus papers;
Wang et al.\ 2025b is YuLan-OneSim, cited only as related work.}
\label{tab:perpaper}
\end{table}

\section{Per-Item Score Distribution}
\label{app:items}

Table~\ref{tab:items} shows the score distribution for Block~D call-site items across the
28 primary call sites. The resolution layer (DR$_1$--DR$_5$) is systematically absent.

\begin{table}[h]
\centering\small
\begin{tabular}{llccccc}
\toprule
Section & Item & Mean & $n(0)$ & $n(1)$ & $n(2)$ & $n$ \\
\midrule
\multirow{2}{*}{Signature}
  & DS\_1 (input bindings)    & 1.81 & 0 & 5  & 22 & 27 \\
  & DS\_2 (output type)       & 1.60 & 0 & 8  & 12 & 20 \\
\midrule
\multirow{7}{*}{Binding}
  & DB\_1 (static text)       & 1.45 & 3 & 6  & 13 & 22 \\
  & DB\_2 (serialization)     & 1.27 & 5 & 6  & 11 & 22 \\
  & DB\_3 (order/placement)   & 0.86 & 5 & 15 & 2  & 22 \\
  & DB\_4 (history policy)    & 1.36 & 2 & 14 & 12 & 28 \\
  & DB\_5 (truncation)        & 1.19 & 6 & 10 & 11 & 27 \\
  & DB\_6 (token budget)      & 0.39 & 17& 11 & 0  & 28 \\
  & DB\_7 (action encoding)   & 1.57 & 3 & 3  & 15 & 21 \\
\midrule
\multirow{5}{*}{Resolution}
  & DR\_1 (parser)            & 0.71 & 11& 14 & 3  & 28 \\
  & DR\_2 (validity)          & 0.75 & 10& 15 & 3  & 28 \\
  & DR\_3 (retry policy)      & 0.15 & 23& 4  & 0  & 27 \\
  & DR\_4 (failure cascade)   & 0.12 & 23& 3  & 0  & 26 \\
  & DR\_5 (failure rate)      & 0.07 & 25& 2  & 0  & 27 \\
\midrule
Delegation & DF\_1 (framework) & 0.53 & 8 & 9  & 0  & 17 \\
\bottomrule
\end{tabular}
\caption{Block~D per-call-site item scores. Items are grouped by section within the
call-site record. The resolution layer is nearly absent from the literature: DR$_3$--DR$_5$
have never scored~2 in any paper.}
\label{tab:items}
\end{table}

\section{Corpus and Exclusions}
\label{app:corpus}

The corpus was drawn from a pre-specified venue set (ACL, EMNLP, NAACL, COLING, NeurIPS,
ICML, ICLR, AAAI, IJCAI, AAMAS, ICWSM, WWW, CHI, JASSS, \emph{Science Advances},
\emph{Nature Human Behaviour}, \emph{PNAS}, arXiv cs.MA/cs.CL/physics.soc-ph), 2023--2026.

\paragraph{Inclusion.} $\geq$1 LLM-driven agent whose behavior is repeatedly elicited over
$\geq$2 timesteps with state carried forward, interacting with at least one other party
(LLM, scripted, or human), in a social, economic, or political scenario.

\paragraph{Flow.} 48 records identified $\to$ 16 excluded at title/abstract screening
$\to$ 32 screened on read $\to$ 4 excluded on read $\to$ 28 coded.

\paragraph{Exclusions on read.} \texttt{horton2023} (E1: one-shot
vignettes, no carried-forward state), \texttt{park2024} (E1: independent questionnaire
responses, no inter-agent interaction), \texttt{abdelnabi2023} (E3: evaluation benchmark,
not social simulation), \texttt{relorder2026} (deferred: retrieval failure).
Screening exclusions: 7$\times$E3 task-completion, 3$\times$E2 non-human actors,
3$\times$E4 methodology papers, 3$\times$E5 framework papers.

\section{Coding Instrument (v1)}
\label{app:instrument}

The full instrument, frozen 2026-08-15 after Pass~1. Every item answers one question:
\emph{Could an independent implementer, reading only this paper and its linked artifacts,
determine this, or would they have to guess?}

\paragraph{Score scale.}
\textbf{2}~(Determined): an implementer could set this correctly without guessing.
\textbf{1}~(Partially determined): named with some content, but at least one behaviorally
relevant aspect remains open; assertion without operational content scores~1, never~2.
\textbf{0}~(Undetermined): absent or constrains nothing.
\textbf{NA}: the system genuinely has no such component.
\textbf{UNVERIFIED}: artifact exists but coder could not obtain it; excluded from
denominators.

\paragraph{Decision-point granularity.}
One decision point = one independently variable choice that could be set differently
without contradicting the text. This is deliberately the same definition as a multiverse
branch point.

\paragraph{Block A: Cognitive Architecture (9 items).}
\begin{enumerate}\setlength{\itemsep}{0pt}
  \item[A1.] Working memory contents per cycle
  \item[A2.] Long-term memory types and backing store
  \item[A3.] Retrieval rule (formula, constants, cardinality)
  \item[A4.] Learning/write rule
  \item[A5.] Internal action space
  \item[A6.] External action space
  \item[A7.] Decision procedure (incl.\ actions per activation)
  \item[A8.] Tools (or NA)
  \item[A9.] Inter-agent channel and visibility parameter
\end{enumerate}

\paragraph{Block B: Identity (6 items).}
\begin{enumerate}\setlength{\itemsep}{0pt}
  \item[B1.] Persona source
  \item[B2.] Persona content
  \item[B3.] Persona surface form (verbatim template)
  \item[B4.] Heterogeneity method
  \item[B5.] Behavioral constraints
  \item[B6.] Identity stability (static/drifting, with update rule)
\end{enumerate}

\paragraph{Block C spine (2 items).}
\begin{enumerate}\setlength{\itemsep}{0pt}
  \item[CS1.] Provider-side non-determinism sources and handling
  \item[CS2.] Execution concurrency and ordering
\end{enumerate}

\paragraph{Block D spine (2 items).}
\begin{enumerate}\setlength{\itemsep}{0pt}
  \item[DG1.] Temporal coupling (step $t$ output $\to$ step $t{+}1$ input)
  \item[DG2.] Scheduling and activation (timestep, who acts, ordering)
\end{enumerate}

\paragraph{Block E: Validation (10 items).}
\begin{enumerate}\setlength{\itemsep}{0pt}
  \item[E1.] Behavioral reference
  \item[E2.] Validation level
  \item[E3a.] Replication count stated
  \item[E3b.] Variance unit
  \item[E4.] Variance reporting
  \item[E5.] Representation sensitivity
  \item[E6.] Population dispersion
  \item[E7a.] Ablation/non-social control
  \item[E7b.] Non-LLM or human baseline
  \item[E8.] Longitudinal stability
\end{enumerate}

\paragraph{Per model role (6 items, repeated).}
Roles: agent, classifier, judge, embedder, generator, external API.
\begin{enumerate}\setlength{\itemsep}{0pt}
  \item[CR1.] Model identity (exact version string)
  \item[CR2.] Build identity (system fingerprint)
  \item[CR3.] Provider, endpoint, API version, access date
  \item[CR4.] Decoding parameters (named unambiguously)
  \item[CR5.] Serving precision and quantization
  \item[CR6.] Training/adaptation (NA if off-the-shelf)
\end{enumerate}

\paragraph{Per call site (15 items, repeated).}
\emph{Signature:}
\begin{enumerate}\setlength{\itemsep}{0pt}
  \item[DS-1.] Input bindings (which state variables feed this call)
  \item[DS-2.] Output type or schema
\end{enumerate}
\emph{Binding:}
\begin{enumerate}\setlength{\itemsep}{0pt}
  \item[DB-1.] Static text (system prompt, verbatim)
  \item[DB-2.] Serialization per slot
  \item[DB-3.] Component order and role placement
  \item[DB-4.] History policy with cardinality
  \item[DB-5.] Truncation/summarization rule with constants
  \item[DB-6.] Token budget and drop order
  \item[DB-7.] Action-space encoding
\end{enumerate}
\emph{Resolution:}
\begin{enumerate}\setlength{\itemsep}{0pt}
  \item[DR-1.] Parser (exact extraction procedure)
  \item[DR-2.] Validity constraints
  \item[DR-3.] Retry policy (count, prompt modification, fallback)
  \item[DR-4.] Failure action and cascade
  \item[DR-5.] Failure rate (reported as result)
\end{enumerate}
\emph{Delegation:}
\begin{enumerate}\setlength{\itemsep}{0pt}
  \item[DF-1.] Third-party framework (library, pinned version, scope)
\end{enumerate}

\section{Worked Example: Block~D for the Naming Game}
\label{app:worked}

Our multiverse arm configurations are themselves Block~D specifications. Below we show
how the baseline configuration (\texttt{base.yaml}) maps to each protocol item.

\begin{verbatim}
experiment:
  population_size: 24
  name_pool: ["M", "Q"]
  memory_length: 5
  payoff_success: 100
  payoff_failure: -50

model:
  provider: "bedrock"
  name: "us.anthropic.claude-sonnet-4-6"
  temperature: 0.7
  top_k: 40
  max_tokens: 1024

branch:
  memory_format: "table"       # DB-2
  memory_order: "recent_first" # DB-3
  score_semantics: "window_sum"# DB-4
  person_framing: "third_person_observer" # DB-1
  pairing: "random_pair_per_step"         # DG2
  parse_failure: "resample_max_3"         # DR-3

fixed:
  randomize_name_order: true   # DB-7
  cot: true                    # DB-1
\end{verbatim}

\paragraph{Item mapping.}
\textbf{DG1} (temporal coupling): step $t$ output (chosen name, payoff) enters step
$t{+}1$ as a memory row. \textbf{DG2}: random-pair-per-step, sequential within step.
\textbf{DS-1}: $f(\text{persona}, \text{memory}[H], \text{score}, \text{name\_pool})$.
\textbf{DS-2}: single letter from name pool.
\textbf{DB-1}: fixed system prompt with third-person observer framing and CoT instruction
(verbatim below).
\textbf{DB-2}: table format (columns: round, Player~1, Player~2, payoff).
\textbf{DB-3}: system prompt $\to$ memory table $\to$ query; recent-first within table.
\textbf{DB-4}: last $H{=}5$ interactions, no retrieval beyond window.
\textbf{DB-5}: hard window truncation at $H$.
\textbf{DB-6}: max\_tokens 1024; no drop order needed (prompt $\ll$ budget).
\textbf{DB-7}: name pool presented as list; order randomized per call.
\textbf{DR-1}: case-insensitive regex \verb|CHOICE:\s*(\S+)|, quotes stripped, matched
against the name pool; if that fails, the last line of the output that equals a pool name
(exact code below).
\textbf{DR-2}: output must be in name pool.
\textbf{DR-3}: resample up to 3 times with identical prompt; uniform random on final failure.
\textbf{DR-4}: uniform random choice; other agents observe a valid name (no null actions).
\textbf{DR-5}: 0.4\% (127/32{,}000 calls across all arms).
\textbf{DF-1}: no third-party framework; custom implementation.

\paragraph{Verbatim prompts (baseline arm).}
The system prompt is three blocks joined by blank lines; \texttt{\{pool\}} is the name pool
in a per-call random order (DB-7), and the memory table holds the last $H{=}5$ interactions,
most recent first. With an empty memory, the table block is replaced by \emph{``Player~1
has no interaction history yet. This is their first interaction.''}

\begin{verbatim}
[system]
You are observing a coordination game between two players. In each round,
both players simultaneously choose a name from the pool: {pool}. If both
players choose the same name, each earns +100 points. If they choose
different names, each loses 50 points. Player 1's objective is to maximize
their own accumulated point tally, conditional on the behavior of their
co-player. You will predict what Player 1 will choose next based on their
history.

Player 1's interaction history (most recent first):

| Round | Player 1's Choice | Co-player's Choice | Outcome | Payoff |
|-------|-------------|------------------|---------|--------|
| {round} | {own} | {partner} | {Match|Mismatch} | {+100|-50} |
...

Player 1's accumulated score: {window_sum}

Think step by step, then predict Player 1's next choice. Keep your
reasoning brief (2-3 sentences).

You MUST end your response with exactly this line:
CHOICE: <name>

where <name> is one of the names from the pool.

[user]
What name will Player 1 choose in the next interaction?
\end{verbatim}

The first-person arm (M4) replaces the observer wording with second-person wording
(``You are playing a coordination game with another player\ldots'', ``What name do you
choose for this interaction?''); the other arms change only the memory block.

\paragraph{Parser and retry (DR-1 to DR-4).}
\begin{verbatim}
def parse_choice(raw_output, name_pool):
    m = re.search(r"CHOICE:\s*(\S+)", raw_output, re.IGNORECASE)
    if m:
        cand = m.group(1).strip().strip('"').strip("'")
        for name in name_pool:
            if cand.upper() == name.upper():
                return name
    for line in reversed(raw_output.strip().split("\n")):
        for name in name_pool:
            if line.strip().upper() == name.upper():
                return name
    return None   # parse failure

# DR-3: on None, resample with the identical prompt up to 3 times;
# DR-4: after the final failure, choose uniformly at random from the pool.
\end{verbatim}

\paragraph{Block~C mapping (per-role: agent).}
\textbf{CR1}: \texttt{us.anthropic.claude-sonnet-4-6} (dated model ID, not a rotating alias).
\textbf{CR2}: not available (Bedrock does not expose system fingerprints).
\textbf{CR3}: AWS Bedrock \texttt{us-east-1}, accessed 2026-07-20 through 2026-08-10.
\textbf{CR4}: temperature 0.7, top\_k 40, max\_tokens 1024; 20 seeds per arm
(\texttt{seed\_offset} 0--19).
\textbf{CR5}: provider-managed (Bedrock); precision not user-configurable.
\textbf{CR6}: not applicable (no fine-tuning or adaptation).

\end{document}